\documentclass[twocolumn]{aastex7}

\usepackage{enumitem}
\usepackage{CJK}

\defcitealias{2025ApJ...991..170Z}{Z25}

\newcommand\be{\begin{eqnarray}}
\newcommand\ee{\end{eqnarray}}

\shorttitle{Hot halo of MW-like galaxies in TNG50}
\shortauthors{Zhang et al.}
\begin{document}
\begin{CJK*}{UTF8}{gbsn}

\title{Probing the Hot Gaseous Halos of Milky Way-like Galaxies in the TNG50 simulation}

\correspondingauthor{Xiaoxia Zhang; Taotao Fang; Hui Li}
\email{zhangxx@xmu.edu.cn; fangt@xmu.edu.cn; hliastro@tsinghua.edu.cn}

\author[0000-0002-8552-2558]{Zhijie Zhang (张志杰)}
\affiliation{Department of Astronomy, Xiamen University, Xiamen, Fujian 361005, People's Republic of China}
\affiliation{School of Physics and Astronomy, Shanghai Jiao Tong University, Dongchuan Road 800, Shanghai 200240, People's Republic of China}
\email{}
\author[0000-0003-4832-9422]{Xiaoxia Zhang (张小霞)}
\affiliation{Department of Astronomy, Xiamen University, Xiamen, Fujian 361005, People's Republic of China}
\email[]{zhangxx@xmu.edu.cn}  
\author[0000-0002-2853-3808]{Taotao Fang (方陶陶)}
\affiliation{Department of Astronomy, Xiamen University, Xiamen, Fujian 361005, People's Republic of China}
\email[]{fangt@xmu.edu.cn}  
\author[0000-0002-1253-2763]{Hui Li (李辉)}
\affiliation{Department of Astronomy, Tsinghua University, Beijing 100084, People's Republic of China}
\email[]{hliastro@tsinghua.edu.cn}  
\author[0000-0003-2630-9228]{Greg L. Bryan}
\affiliation{Department of Astronomy, Columbia University, 550 West 120th Street, New York, NY 10027, USA}
\email{}
\author[0000-0003-3816-7028]{Federico Marinacci}
\affiliation{Department of Physics and Astronomy "Augusto Righi", University of Bologna, via Gobetti 93/2, I-40129 Bologna, Italy}
\affiliation{INAF, Astrophysics and Space Science Observatory Bologna, Via P. Gobetti 93/3, I-40129 Bologna, Italy}
\email{}
\author[0000-0002-5653-0786]{Paul Torrey}
\affiliation{Department of Astronomy, University of Virginia, 530 McCormick Road, Charlottesville, VA 22903, USA}
\email{}
\author[0000-0001-8593-7692]{Mark Vogelsberger}
\affiliation{Department of Physics and Kavli Institute for Astrophysics and Space Research, Massachusetts Institute of Technology, Cambridge, MA 02139, USA}
\email{}
\author[0000-0003-4874-0369]{Junfeng Wang (王俊峰)}
\affiliation{Department of Astronomy, Xiamen University, Xiamen, Fujian 361005, People's Republic of China}
\email{}
\author[0000-0001-9603-7673]{Haiguang Xu (徐海光)}
\affiliation{School of Physics and Astronomy, Shanghai Jiao Tong University, Dongchuan Road 800, Shanghai 200240, People's Republic of China}
\email{}
\author[0000-0003-3230-3981]{Qingzheng Yu (余清正)}
\affiliation{Dipartimento di Fisica e Astronomia, Universit\`a degli Studi di Firenze, Via G. Sansone 1, 50019 Sesto Fiorentino, Firenze, Italy}
\affiliation{INAF - Osservatorio Astrofisico di Arcetri, Largo E. Fermi 5, I-50125 Firenze, Italy}
\email{}
\author[0000-0003-3564-6437]{Feng Yuan (袁峰)}
\affiliation{Center for Astronomy and Astrophysics and Department of Physics, Fudan University, Shanghai 200438, People's Republic of China}
\email{}

\begin{abstract}
The origin and structure of the hot ($T\gtrsim10^6$\,K) gaseous halo around Milky Way (MW)-mass galaxies provide a critical test for galaxy formation models. We perform a comprehensive comparison for a sample of MW analogues from the TNG50 cosmological simulation by generating synthetic soft X-ray emission and \ion{O}{7}/\ion{O}{8} absorption lines, viewed from both internal (Solar) and external perspectives. The simulated halos successfully reproduce the observed global soft X-ray luminosity, inner-halo X-ray surface brightness, emission measure, and \ion{O}{7} absorption strength. However, two interconnected discrepancies are identified. First, the azimuthally averaged X-ray surface brightness profile from external viewpoints declines too steeply with radius compared to the extended emission detected in eROSITA stacking of SDSS galaxies, falling below the observations by up to $\sim 1$\,dex at $R \gtrsim 100$\,kpc. Second, the halos systematically underproduce \ion{O}{8} absorption, with a median equivalent width $\sim 65\%$ lower than that observed in the Galactic halo, pointing to a deficit of hotter-phase gas at $T\sim(1.6-3.2)\times10^6$\,K. These findings indicate that the simulated hot halos are too spatially compact and lack a hotter gas phase, suggesting that the TNG50 feedback model, while generating hot gas, deposits energy too centrally and too vigorously to sustain a gently extended, multi-phase corona.
\end{abstract}


\section{Introduction}

The presence of an extended, hot gaseous halo is a well-established theoretical prediction for Milky Way (MW)-mass galaxies. In the standard galaxy formation paradigm, gas accreted from the intergalactic medium is shock-heated to temperatures of several million Kelvin, forming a diffuse corona that may constitute a major baryonic component of the halo \citep[][]{1956ApJ...124...20S,1978MNRAS.183..341W,1991ApJ...379...52W, 2017ARA&A..55..389T}. The thermodynamic state and spatial distribution of this corona are governed by the competing effects of gravitational heating, radiative cooling, and energetic feedback from stars and active galactic nuclei (AGN), as well as cosmic rays \citep[e.g.,][]{2025OJAp....8E..78H}. A key question, therefore, is the relative importance of these mechanisms in shaping the hot halo.

Observations provide compelling evidence for hot gaseous halos around MW-like galaxies. Within our own Galaxy,  detections of both the diffuse soft X-ray background \citep[e.g.,][]{1997ApJ...485..125S, 2009PASJ...61..805Y, 2013ApJ...773...92H} and absorption/emission line from highly ionized species like \ion{O}{7} and \ion{O}{8} \citep[e.g.,][]{2003ApJ...586L..49F, 2006ApJ...644..174F, 2012ApJ...756L...8G, 2015ApJS..217...21F, 2015ApJ...800...14M, 2018ApJS..235...28L, 2024ApJS..271...62P} confirm the existence of million-degree plasma extending into the halo. Studies of nearby, edge-on galaxies revealing extraplanar X-ray emission morphologically associated to star-forming disks and superbubbles \citep[e.g.,][]{2004ApJ...606..829S, 2006A&A...448...43T, 2013MNRAS.428.2085L}. On larger scales, stacking analyses of galaxy samples have statistically detected extended, low-surface-brightness X-ray emission around MW-mass galaxies, suggesting the presence of substantial coronae on scales of several tens to hundreds of kiloparsecs \citep[e.g.,][]{2022ApJ...936L..15C, 2022A&A...666A.156C, 2025A&A...693A.197Z, 2026arXiv260116499H}. 

Numerical simulations have yielded divergent conclusions regarding the origin of the hot halo \citep{2020NatRP...2...42V}. Some models suggest stellar feedback can generate significant X-ray emission, though often more centrally concentrated than observed \citep[e.g.,][]{2015ApJ...800..102H, 2015ApJ...813L..27P, 2020ApJ...898..148L}, while others emphasize the critical role of AGN feedback in regulating the large-scale thermodynamic state of the circumgalactic medium \citep[CGM;][]{2020MNRAS.494..549T, 2020ApJ...893L..24O}. 
Recent X-ray predictions from TNG, EAGLE, and SIMBA have revealed substantial diversity in hot CGM properties \citep[e.g.,][]{2023MNRAS.525.1976T, 2024ApJ...969...85S}, underscoring the need for detailed comparisons with new observational data. 
Our previous work with the high-resolution SMUGGLE model \citep[starting without a pre-existing hot corona;][]{2019MNRAS.489.4233M, 2020MNRAS.499.5862L} showed that stellar feedback alone can reproduce the warm Galactic halo traced by \ion{O}{6} \citep{2024ApJ...962...15Z}, but fails to generate the extended hot corona seen in X-ray observations, with emission confined to the inner regions \citep[][hereafter Z25]{2025ApJ...991..170Z}. This highlights the necessity for a cosmological framework that self-consistently includes both cosmological gas accretion and AGN feedback to advance our understanding of the hot halo's origin \citep{2013ApJ...772...97B}.

As the highest-resolution realization of the IllustrisTNG project, the TNG50 simulation provides a suitable framework for studying galaxy formation in a full cosmological context \citep{2013MNRAS.436.3031V, 2014Natur.509..177V}. It self-consistently includes cosmological gas accretion and coupled stellar and AGN feedback \citep{2019MNRAS.490.3234N, 2019MNRAS.490.3196P}, the key processes expected to shape the hot halo. Moreover, its combination of a cosmologically representative volume and high mass resolution makes TNG50 well-suited for studying the detailed properties of the hot gaseous corona around MW-mass galaxies.

In this work, we generate synthetic observations of the hot gaseous halo from a sample of MW-analogues in TNG50, focusing on soft X-ray emission and \ion{O}{7} and \ion{O}{8} absorption lines from both internal (Solar) and external viewpoints. Our analysis aims to determine whether this state-of-the-art simulation can reproduce the observed properties of hot halo gas.MW-mass galaxies reside in a mass regime where stellar and AGN feedback are comparably important, making the hot halo a sensitive probe of the AGN feedback implementation and its interplay with stellar feedback. A detailed comparison with observations thus provides a critical test of the AGN feedback model in TNG50, and can help identify specific limitations in the current implementation.

The paper is structured as follows. Section~\ref{sec:method} describes the TNG50 simulation, the selection criteria for our MW-like galaxy sample, and our methodology for generating synthetic X-ray and absorption-line observables. We present our main results in Section~\ref{sec:result}, comparing the simulated hot halo properties, such as X-ray luminosities, surface brightness profiles, emission measures, and \ion{O}{7} and \ion{O}{8} absorption strengths, with a suite of multi-scale observations. The implications of these findings for models of hot halo origin are discussed in Section~\ref{sec:discuss}. We summarize our conclusions in Section~\ref{sec:summary}.

%%%%%%%%%%%%%%%%%%%%%%%%%%%%%%%%%%%%%%%%%%%%%%%%%%%%%%%%%%%%%
\section{Methodology}
\label{sec:method}

\subsection{The TNG50 Simulations}
\label{sec:Simulations}

The TNG50 simulation is the highest-resolution run of the IllustrisTNG project \citep{2018MNRAS.480.5113M, 2018MNRAS.477.1206N, 2018MNRAS.475..624N, 2018MNRAS.475..648P, 2018MNRAS.475..676S}. It is designed to model galaxy formation within a full cosmological context \citep{2019MNRAS.490.3234N, 2019MNRAS.490.3196P}. The simulation builds upon and substantially improves its predecessor, the original Illustris simulation \citep{2014MNRAS.444.1518V}, through key advancements including the inclusion of magnetohydrodynamics and a comprehensively revised model for feedback from stars and AGN \citep{2017MNRAS.465.3291W, 2018MNRAS.473.4077P}. These improvements, implemented in the moving-mesh code \textsc{AREPO} \citep{2010MNRAS.401..791S}, enable TNG50 to produce a galaxy population in better agreement with a range of observations \citep[e.g.,][]{2019MNRAS.490.3234N}. 

The simulation tracks structure formation in a periodic volume with a comoving side length of approximately 51.7\,Mpc. The high resolution is reflected in the mass resolution of $8.5 \times 10^{4}\,M_\odot$ for baryonic particles and $4\times10^5\,M_\odot$ for dark matter particles. The spatial resolution of the gas is determined by an adaptive mesh that can refine cells down to scales of several tens of parsecs.  At $z=0$, the gravitational softening lengths are 0.29\,kpc and 0.74\,kpc (proper) for stellar and dark matter particles, respectively.

For our study, a crucial aspect is that the TNG model self-consistently includes the cosmological accretion of gas and the coupled effects of stellar and AGN feedback. AGN feedback operates in two distinct modes: a thermally dominated ``quasar mode'' at high accretion rates, and a kinetically dominated ``radio mode'' at low accretion rates \citep{2017MNRAS.465.3291W}. 
The kinetic ``radio mode'' is active at late times when the black hole accretion rate drops sufficiently. In this mode, AGN feedback injects energy and momentum into the surrounding gas, driving outflows that can displace central hot gas to larger radii, thereby reducing the central gas density and suppressing radiative cooling. This mechanism is intended to help maintain an extended, diffuse corona.

%%%%%%%%%%%%%%%%%%%%%%%%%%%%%%%%%%%%%%%%%%%%%%%%%%%%%%%%%%%%%
\subsection{The MW-like Galaxy Sample}
\label{sec:sample}

Our sample of MW analogues is constructed from the parent sample of disky central galaxies at $z=0$ in the TNG50 simulation \citep{2023MNRAS.518.5754R, 2024MNRAS.535.1721P}. The parent sample comprises galaxies with stellar masses in the range $10^{10.5} < M_*/ M_\odot < 10^{10.9}$, measured within a 30\,kpc aperture.
Furthermore, these galaxies reside in halos with masses $M_{\mathrm{200, c}} < 10^{13}\,M_\odot$ (with typical values of $10^{11.7}-10^{12.5}\,M_\odot$), and satisfy an isolation criterion to exclude those in rich group environments.

\begin{figure}[htbp]
\centering
\includegraphics[width=0.47\textwidth]{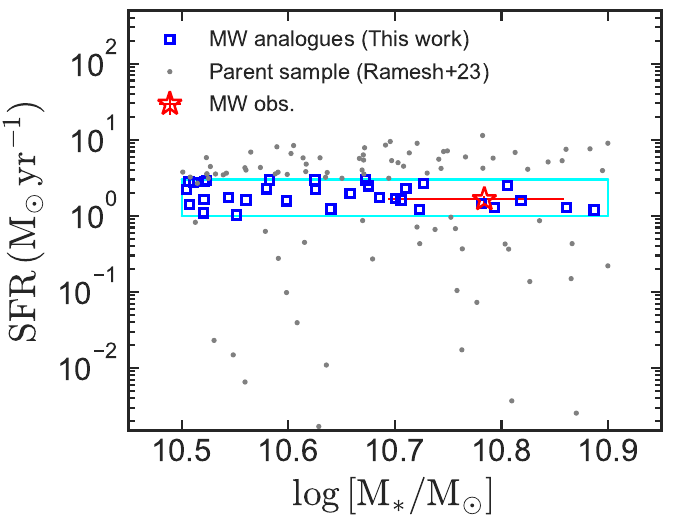}
\caption{Selection of the MW analogues in the stellar mass--SFR plane. Gray dots show the parent population of MW-mass galaxies in TNG50 \citep{2023MNRAS.518.5754R}. The cyan rectangle outlines our selection criteria in stellar mass and SFR. Blue squares highlight the 32 galaxies meeting these criteria, which form our final sample. The red star marks the observed values of the MW \citep{2015ApJ...806...96L}.
\label{fig-1}}
\end{figure}

To match the present-day evolutionary state of the MW, we impose an additional constraint on the instantaneous star formation rate (SFR), selecting galaxies with $1<\rm{SFR}/(M_\odot\,{\rm yr^{-1}}) <3$. This range is consistent with modern estimates for the MW \citep[the red star in Figure~\ref{fig-1};][]{2015ApJ...806...96L} and ensures that our sample galaxies have a comparable level of recent stellar feedback. Applying this SFR criterion to the parent sample yields our final set of 32 MW analogues, shown as the blue squares in Figure~\ref{fig-1}.

%%%%%%%%%%%%%%%%%%%%%%%%%%%%%%%%%%%%%%%%%%%%%%%%%%%%%%%%%%%%%
\subsection{Synthetic Observables}
\label{sec:synthetic_observations}

To compare the simulated hot halo gas with available observations, we generate synthetic data for two primary diagnostics: soft X-ray (0.5--2.0\,keV) emission and absorption lines from highly ionized species. The calculation of soft X-ray emission follows the same procedure as \citetalias{2025ApJ...991..170Z} and is summarized below.

\subsubsection{X-ray Emission}
\label{sec:xray_method}

The emissivity for each gas cell under the assumption of collisional ionization equilibrium is
\be
\epsilon_{\mathrm{X}} = n_e n_{\mathrm{H}} \, \Lambda(T, Z),
\label{eq:emissivity}
\ee
where \(n_e\) and \(n_{\mathrm{H}}\) are the electron and hydrogen number densities, \(T\) and \(Z\) are the gas temperature and metallicity, and \(\Lambda(T, Z)\) is the cooling function calculated using the APEC model \citep[v3.0.9;][]{2001ApJ...556L..91S}.\footnote{Star-forming gas cells are not explicitly removed, since their low temperatures ($T \ll 10^6$ K) make their contribution to the soft X-ray band negligible; similarly, wind particles are not treated separately, as they have cooled and mixed with the ambient halo gas by $z=0$.}

To generate observables comparable to actual observations, we project quantities along specific lines of sight, defining two primary viewpoints. For the \textbf{internal (Solar) viewpoint}, we first determine the galactic disk plane by computing the net angular momentum vector of star particles within a radial shell of (1--2)\(R_{*,\mathrm{half}}\) to robustly trace the large-scale stellar disk, where \(R_{*,\mathrm{half}}\) is the three-dimensional stellar half-mass radius. An observer is then placed at 8.2\,kpc on this plane \citep{2016ARA&A..54..529B}.

To obtain robust statistics and mitigate potential bias from a single fixed location, we adopt a multi-observer approach commonly used in synthetic observations \citep[e.g.,][]{2020ApJ...896..143Z, 2024ApJ...962...15Z}. We place four such observers in the disk plane, separated by \(90^\circ\) in azimuth. From each observer, we generate 500 random sightlines. The Galactic longitude \(l\) is sampled uniformly over \([0^\circ, 360^\circ)\), and the latitude \(b\) is sampled over \([-90^\circ, 90^\circ]\) with a probability density proportional to \(\cos(b)\). When comparing with MW halo observations that typically avoid the disk \citep{2013ApJ...773...92H}, we restrict our analysis to sightlines with \(|b| > 30^\circ\).
As a test of potential resolution effects, we exclude gas within 3\,kpc of the Solar observers and find that the median of all high-latitude sightline quantities changes by less than 0.1\,dex. We therefore retain all gas in the analysis.

For the \textbf{external viewpoint}, designed for comparison with randomly oriented external galaxies, we define the observation geometry by generating 100 randomly oriented lines of sight for each galaxy \citep[following the common practice in synthetic observations, e.g.,][]{2020ApJ...896..143Z}. These sightlines will be used to construct azimuthally averaged radial surface brightness profiles.

\begin{figure*}[htbp]
\centering
\includegraphics[width=0.9\textwidth]{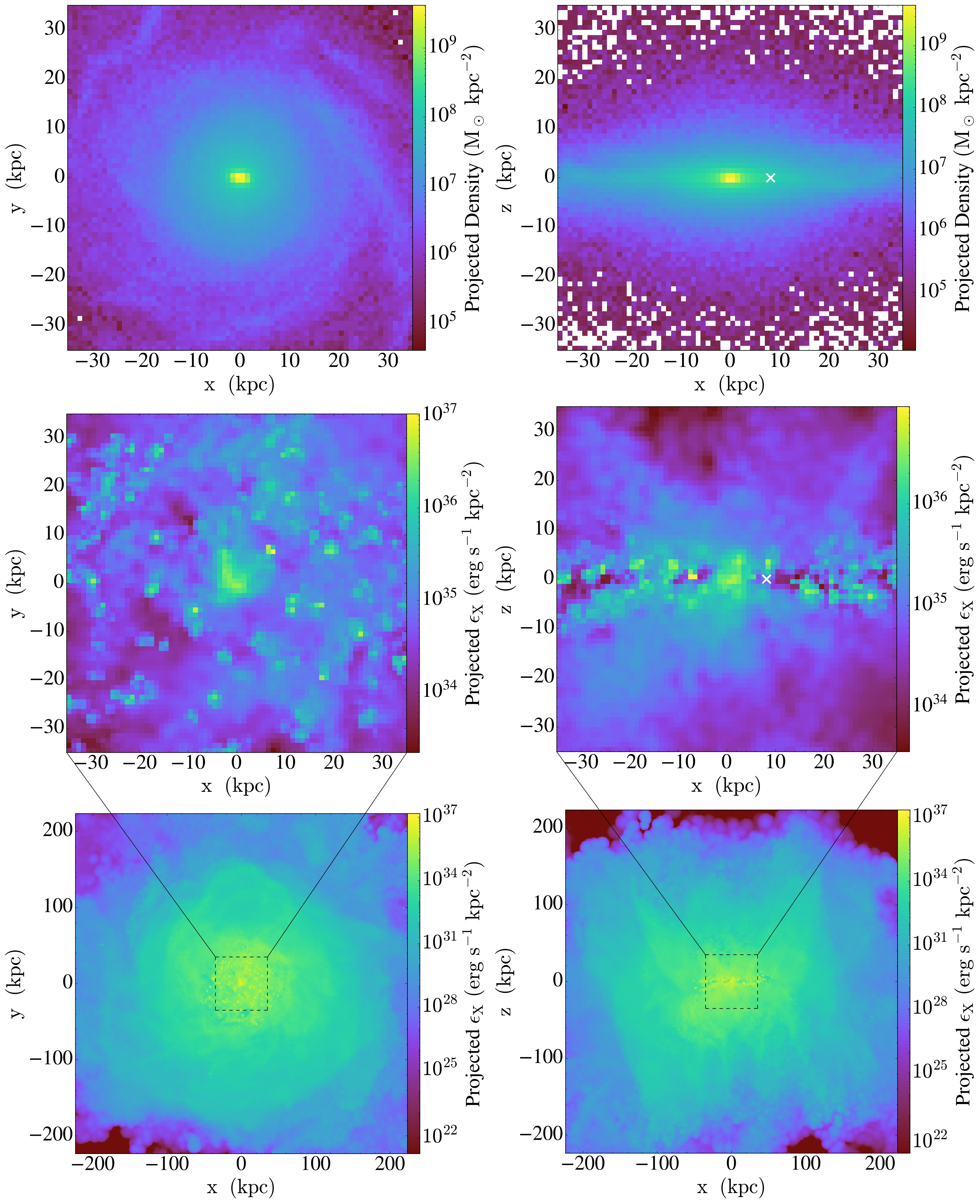}
\caption{Visualization of a representative simulated galaxy (ID = 520885). The top row shows the face-on (left) and edge-on (right) projected stellar mass surface density within $\pm 35$\,kpc. The middle row displays the corresponding soft X-ray (0.5--2.0\,keV) emissivity of the inner region, revealing complex, centrally concentrated hot gas. The bottom row extends the view to the halo scale ($R_{\rm 200,c}$), with dashed boxes outlining the $\pm 35$\,kpc region shown above. A white cross marks the adopted Solar position (8.2\,kpc) in the edge-on views.
\label{fig-2}}
\end{figure*}

Figure~\ref{fig-2} shows a representative TNG50 MW analogue. The galaxy hosts a well-defined stellar disk, as seen in the face-on and edge-on views (top row). The soft X-ray emissivity (middle row) is concentrated in the inner region, reflecting the influence of stellar feedback and inner-halo processes. Extending the projection to the virial radius (bottom row) reveals that the hot halo becomes more diffuse on larger scales but retains noticeable structure.

From the emissivity, we derive several key observables using the {\tt yt} analysis toolkit \citep{2011ApJS..192....9T} to perform the necessary summations, integrations, and projections.
The X-ray luminosity within a specified volume is obtained by direct summation over gas cells:
\be
L_{\rm X} = \sum \epsilon_{\mathrm{X}} \, V_{\mathrm{cell}}.
\label{eq:lum}
\ee
For comparisons with external galaxy samples, we compute \(L_{\rm X}\) within a cylindrical region centered on the galactic center and extending vertically $\pm 5$\,kpc  from the disk midplane. This corresponds to the median extent used for spectral analysis in the observed sample \citep{2016MNRAS.457.1385W}. 

The X-ray surface brightness \(S_{\rm X}\) along a given sightline is computed by integrating the emissivity 
\be
S_{\rm X} = \frac{1}{4\pi} \int \epsilon_{X} \, ds,
\label{eq:sb}
\ee
where $s$ is path length of the sightline.
The factor \(1/(4\pi)\) gives \(S_{\rm X}\) in units of intensity per steradian. For comparison with the Galactic halo observations of Henley et al. (2013), we convert this to \(\mathrm{erg\,s^{-1}\,cm^{-2}\,deg^{-2}}\) when constructing all-sky maps (Figure 5).

For the internal (Solar) viewpoint, the integration is performed from the observer out to a radius of 260\,kpc, matching the typical extent of observational constraints for the Galactic halo \citep[e.g.,][]{2013ApJ...773...92H}. This yields sightline-specific values of $S_{\rm X}$ projected onto a sky defined by Galactic \((l, b)\) coordinates.

For the external viewpoint, we project the emission along each random sightline through the entire simulated halo. 
To enable a fair comparison with the eROSITA stacking profiles, we convolve the projected two-dimensional images with a Gaussian kernel matching the eROSITA PSF before azimuthal averaging. We adopt a full width at half maximum (FWHM) of $30\arcsec$ \citep{2024A&A...682A..34M}, corresponding to $\approx 47$\,kpc at the median redshift of the MW-mass CEN sample (\citet{2024A&A...690A.267Z}; $z \approx 0.08$).
We then compute the azimuthally averaged radial surface brightness profile $S^{\rm ext}_{\rm X}(R)$ by collecting and averaging the $S_{\rm X}$ values from 100 random sightlines for each radial bin relative to the galaxy center.

Beyond the integrated luminosity and projected surface brightness, the thermodynamic state of the hot halo gas provides crucial diagnostics for its origin and heating mechanisms. 
The emission measure (EM) for the hot gas phase is calculated by integrating the density squared for gas cells with \(T > 10^6\,\mathrm{K}\):
\be
\mathrm{EM} = \int n_e n_{\mathrm{H}} \, ds,
\label{eq:em}
\ee
and the emission-weighted temperature (\(T_{\mathrm{ew}}\)) is derived by weighting the gas temperature with the X-ray emissivity:
\be
T_{\mathrm{ew}} = \frac{\int T \, \epsilon_{\mathrm{X}} \, ds}{\int \epsilon_{\mathrm{X}} \, ds}.
\label{eq:tew}
\ee
Because \(\epsilon_{\mathrm{X}}\) is negligible for gas at temperatures much below \(10^6\,\mathrm{K}\) in the soft X-ray band, \(T_{\mathrm{ew}}\) effectively represents the characteristic temperature of the X-ray-emitting hot gas. These quantities are computed for the same sets of sightlines as \(S_{\rm X}\), enabling direct comparison with observational derivations from X-ray spectral fitting.

%%%%%%%%%%%%%%%%%%%%%%%%%%%%%%%%%%%%%%%%%%%%%%%%%%%%%%%%%%% 

\subsubsection{Absorption Lines}
\label{sec:ew_method}

To compare with absorption-line observations of the MW's hot halo, we generate synthetic profiles for the \ion{O}{7} and \ion{O}{8} K\(\alpha\) transitions, which have characteristic wavelengths of approximately 21.6\AA \ and 19.0\,\AA, respectively. These ions serve as key diagnostics for the hot gas. Under the assumption of collisional ionization equilibrium, their fractional abundances peak at characteristic temperatures of \(\sim 8 \times 10^5\)\,K and \(\sim 2.5 \times 10^6\)\,K, respectively \citep{1998A&AS..133..403M}. From the synthetic profiles, we compute the equivalent width (EW) for direct comparison with measurements from the soft X-ray spectra of background AGN.

To model the absorption line, we first compute the number density of the ion in each gas cell. Using \ion{O}{7} as an example:
\be
n(\text{\ion{O}{7}}) = n_{\mathrm{O}} \, f_{\rm O\,{\textsc{vii}}}(T), 
\ee
where \(n_{\mathrm{O}}\) is the oxygen number density in the cell, computed from the tracked oxygen mass fraction in the TNG50 simulation. 
The ionization fraction \(f_{\mathrm{O\,\textsc{vii}}}(T)\) is obtained assuming collisional ionization equilibrium \citep{1998A&AS..133..403M}. An identical calculation is performed for \ion{O}{8} using its corresponding ionization fraction.

\begin{figure}[tbp]
\centering
\includegraphics[width=0.47\textwidth]{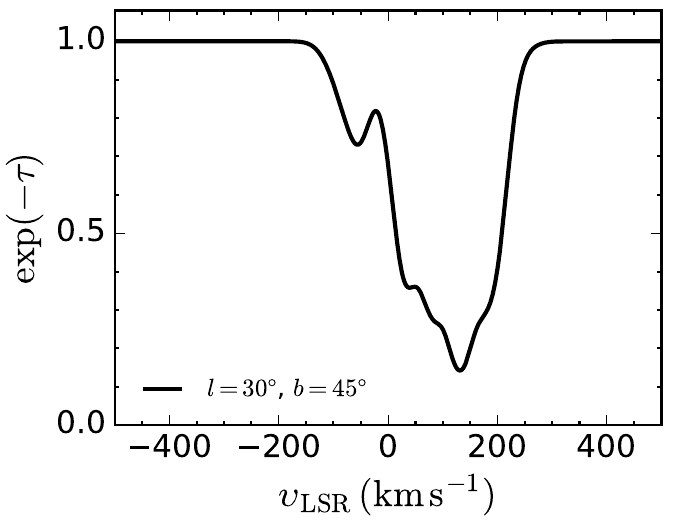}
\caption{Example synthetic \ion{O}{7} K$\alpha$ absorption line profile (ID = 555601) along a sightline with Galactic coordinates $(l, b) = (30^\circ, 45^\circ)$. The optical depth $\tau$ is plotted against the Local Standard of Rest (LSR) velocity. 
\label{fig-3}}
\end{figure}

For a given sightline, the optical depth profile \(\tau(\lambda)\) is determined by the integrated column density of the ion, its oscillator strength, and the line-of-sight velocity structure of the gas. We incorporate line broadening from thermal motions and shifts in the line center arising from both Hubble expansion and the peculiar velocities of the intervening gas \citep{2002ApJ...564..604F}. 
To isolate absorption features arising from the Galactic hot halo, consistent with observational analyses \citep{2018ApJS..235...28L}, we only include gas with a line-of-sight velocity $|v_{\rm LSR}| \lesssim 500\,{\rm km\,s^{-1}}$ relative to the Local Standard of Rest. 
This cut excludes high-velocity clouds and satellite features, which contribute negligibly along high-latitude sightlines. Figure~\ref{fig-3} shows an example of the resulting \ion{O}{7} optical depth profile.

Finally, the equivalent width is computed by integrating the resulting absorption profile over the wavelength,
\be
\mathrm{EW} = \int \left( 1 - e^{-\tau(\lambda)} \right) d\lambda,
\label{eq:ew}
\ee
and we apply this method to sightlines originating from the Solar position defined in Section~\ref{sec:xray_method}, probing the halo in various Galactic $(l, b)$ directions while excluding $b<20^\circ$. This approach directly mimics the observational strategy of using background AGN to study the MW's hot gaseous halo \citep{2015ApJS..217...21F}.

%%%%%%%%%%%%%%%%%%%%%%%%%%%%%%%%%%%%%%%%%%%%%%%%%%%%%%%%%%%
\section{Results} 
\label{sec:result}

\subsection{X-ray Luminosity and Surface Brightness}
\label{sec:luminosity}

\begin{figure}[tbp]
\centering
\includegraphics[width=0.47\textwidth]{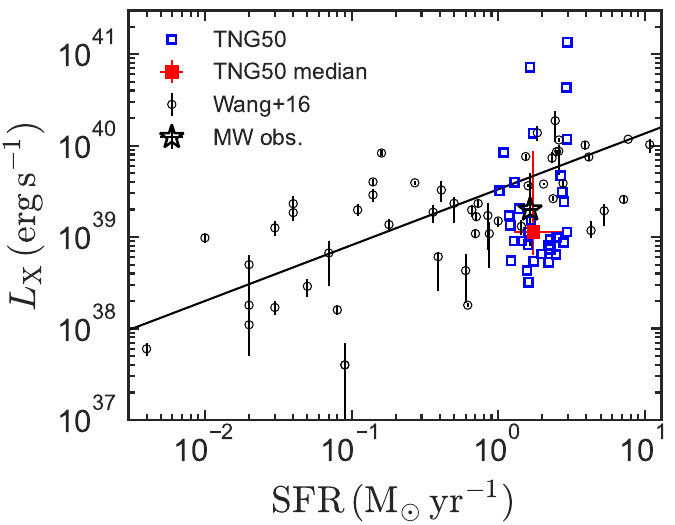}
\caption{Soft X-ray luminosity versus star formation rate for simulated MW analogues. The simulated data are shown as blue squares, with the median and $1\sigma$ scatter of the sample indicated by the red square and error bars. The empirical relation for nearby, highly inclined disk galaxies \citep[][]{2016MNRAS.457.1385W} is shown for comparison, together with the observed values of the MW \citep{1997ApJ...485..125S, 2015ApJ...800...14M}.
\label{fig-relation}}
\end{figure}

\begin{figure*}[tbp]
\centering
\includegraphics[width=0.9\textwidth]{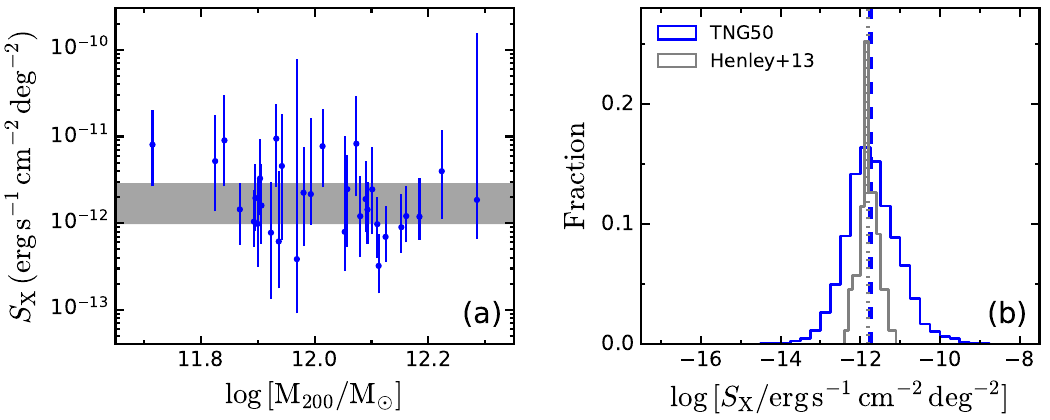}
\caption{(a) Median surface brightness and $1\sigma$ scatter from the Solar perspective, restricted to sightlines with $|b|>30^{\circ}$, plotted against halo mass ($M_{200}$). The gray band shows the observed range for the Galactic halo \citep{2013ApJ...773...92H}. (b) Surface brightness distribution for all individual Solar sightlines in the TNG50 sample (blue histogram) compared with the same observed data (gray histogram). Vertical lines mark the respective median values.
\label{fig-sb}}
\end{figure*} 

Figure~\ref{fig-relation} compares the soft X-ray luminosities of our TNG50 MW analogues with the empirical $L_{\rm X}$--SFR relation established for nearby, highly inclined disk galaxies \citep[][]{2016MNRAS.457.1385W}. While the SFRs of our selected galaxies are narrowly distributed around the MW's value, their X-ray luminosities span nearly three orders of magnitude. The median luminosity of the TNG50 sample agrees with both the observed value for the MW \citep[][]{1997ApJ...485..125S, 2015ApJ...800...14M} and the empirical trend defined by external disk galaxies. This indicates that the simulation produces a characteristic X-ray luminosity for MW-like galaxies that is consistent with current observational constraints. The large predicted scatter suggests a greater halo-to-halo diversity than is currently observed.

Figure~\ref{fig-sb} examines the soft X-ray surface brightness from the Solar viewpoint. Panel (a) shows the median $S_{\rm X}$ for each TNG50 MW analogue, compared to the observed range for the Galactic halo \citep[gray band;][]{2013ApJ...773...92H}. The spread in median values across the simulated sample is approximately three times wider than the observed range, although most galaxies still fall within or near the observational band, indicating that the characteristic surface brightness is consistent with MW measurements.

Figure~\ref{fig-sb}(b) compares the distribution of $S_{\rm X}$ values from individual Solar sightlines with the observational distribution. The simulated sightlines span over four orders of magnitude, significantly broader than the observed distribution, reflecting the highly inhomogeneous and centrally concentrated hot gas in the simulations (see also Figure~\ref{fig-2}). The narrower observed distribution may partly result from the sensitivity limit of the observations, which truncates the low-surface-brightness end, whereas the simulation includes arbitrarily faint sightlines. Applying a realistic detection threshold to the simulations would likely narrow the simulated distribution. A detailed investigation of this effect is beyond the scope of this paper. We note, however, that individual halos can reproduce the observed scatter: Halo~569251, which best matches the \citet{2013ApJ...773...92H} data, has a median $S_{\rm X}$ of $1.44^{+1.49}_{-0.87} \times 10^{-12}$\,erg\,s$^{-1}$\,cm$^{-2}$\,deg$^{-2}$, consistent with the observed value of $1.54^{+1.35}_{-0.56} \times 10^{-12}$\,erg\,s$^{-1}$\,cm$^{-2}$\,deg$^{-2}$.

\begin{figure}[htbp]
\centering
\includegraphics[width=0.47\textwidth]{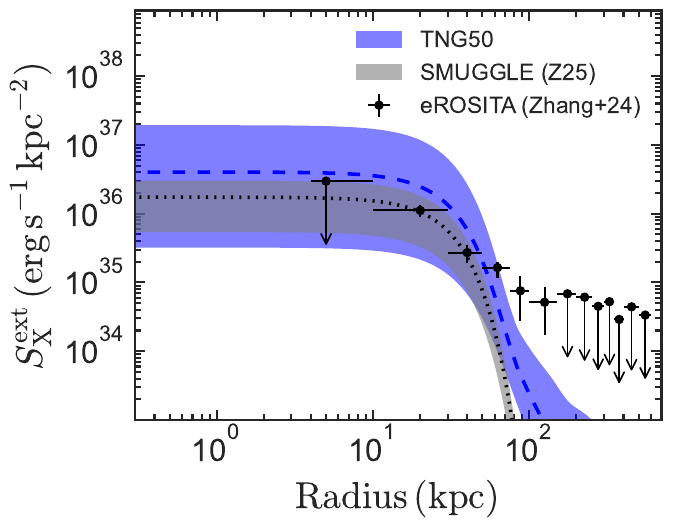}
\caption{Azimuthally averaged soft X-ray surface brightness profile as a function of projected radius. The blue band and dashed line show the TNG50 profiles (mean and $1\sigma$ scatter) after convolution with the eROSITA PSF. Black points with error bars are the hot CGM profile of the MW-mass CEN sample from \citet{2024A&A...690A.267Z}. The gray band and dotted line show the SMUGGLE simulation of an isolated MW-type galaxy \citepalias{2025ApJ...991..170Z}, also convolved with the eROSITA PSF for consistency.
\label{fig-sxp}}
\end{figure} 

To characterize the global spatial distribution, Figure~\ref{fig-sxp} presents the azimuthally averaged radial surface brightness profile from an external viewpoint. The simulated profile declines with increasing projected radius, exhibiting a pronounced steepening beyond $r \gtrsim 30$\,kpc. The observational comparison is based on the eROSITA stacking analysis of the MW-mass CEN sample from \citet{2024A&A...690A.267Z}, which covers $\log (M_*/M_\odot) = 10.5$--$11.0$ and consists of central galaxies at $z \approx 0.02$--$0.10$ (median $z \approx 0.08$) selected using a halo-based group finder to minimize satellite contamination. Our sample is also restricted to isolated centrals, making the comparison reasonable. For a consistent comparison, we convolve the simulated profiles with the eROSITA PSF before azimuthal averaging (see Section~\ref{sec:xray_method}). Compared to the SDSS stacking result, the TNG50 model systematically underestimates the surface brightness at large radii: the convolved profiles are broadly consistent with the observations within $R \sim 50$\,kpc, but fall below them by up to $\sim 1$\,dex at $R \gtrsim 100$\,kpc.

We note, however, that the nearby $L^*$ galaxy sample studied by \citet{2026arXiv260116499H} exhibits a steeper X-ray surface brightness profile than the SDSS sample, and their comparison with TNG50 also shows broad consistency. Their inner profile shows a rising trend that our PSF-convolved profiles do not capture, likely owing to the larger physical PSF scale ($\sim 47$\,kpc at $z\sim 0.08$) adopted for the SDSS comparison. 
The SMUGGLE simulation of an isolated MW-type galaxy \citepalias{2025ApJ...991..170Z}, also convolved with the eROSITA PSF for consistency, is shown for reference; its mean profile falls below the TNG50 predictions at all radii but lies within the TNG50 scatter within $R \sim 70$\,kpc, while dropping below it at larger radii. Despite the improvement over stellar-feedback-only models, the TNG50 halos remain more centrally concentrated than the SDSS sample at large radii, indicating a deficit of diffuse X-ray emission relative to this population.

 %%%%%%%%%%%%%%%%%%%%%%%%%%%%%%%%%%%%%%%%%%%%%%%%%%%%%%%%%%% 
 \subsection{Emission Measure and Temperature}
 \label{sec:em}

\begin{figure*}[tbp]
\centering
\includegraphics[width=0.9\textwidth]{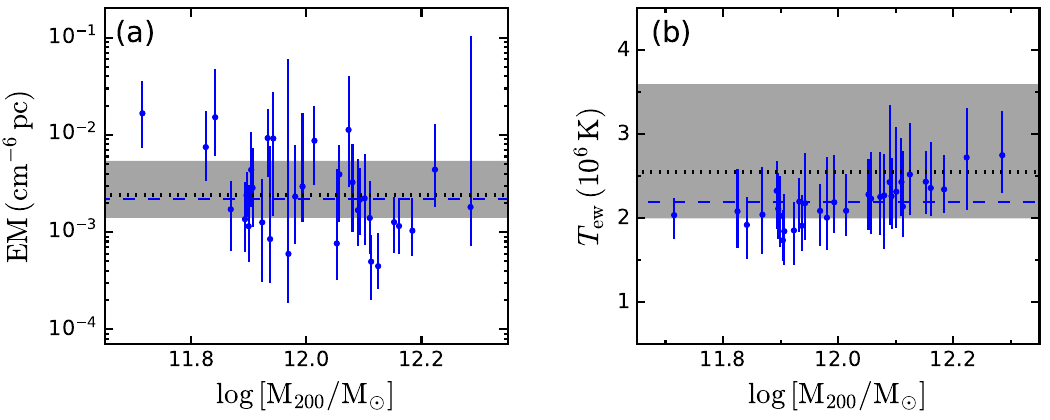} 
\caption{Emission measure (left) and emission-weighted temperature (right) for MW-mass galaxies from TNG50 as a function of halo mass ($M_{200}$). Error bars represent the $1\sigma$ scatter among sightlines with $|b|>30^\circ$. The grey band marks the observed $1\sigma$ range for the MW's hot halo from combined {\it XMM-Newton} and {\it Suzaku} observations \citep[][]{2013ApJ...773...92H, 2018ApJ...862...34N}. The blue dashed and grey dotted lines indicate the median values of the simulated sample and the observational range, respectively.
\label{fig:em_temp}}
\end{figure*}

Decomposing the X-ray surface brightness into emission measure (density-sensitive) and emission-weighted temperature helps to identify whether differences between simulations and observations stem from the gas density distribution or the thermal state of the hot halo. We analyze the emission measure and emission-weighted temperature along individual sightlines from the Solar viewpoint, as defined in Eqs.~(\ref{eq:em}) and (\ref{eq:tew}). 

Figure~\ref{fig:em_temp}(a) presents the EM for each MW analogue. The simulated galaxies exhibit a large scatter in halo-averaged EM. Despite this large galaxy-to-galaxy variation, the median EM of the sample agrees well with the observational estimate for the Galactic halo, indicating that TNG50 produces a hot gas density distribution consistent with that inferred for the MW.
Figure~\ref{fig:em_temp}(b) shows the emission-weighted temperature ($T_{\rm ew}$). A weak positive correlation with $M_{200}$ is observed, reflecting the underlying gravitational potential.  For most galaxies, the simulated temperatures fall within the observational constraints for the MW's hot halo, suggesting that the balance between gravitational heating, cooling, and feedback in TNG50 yields realistic characteristic temperatures.

\begin{figure*}[tbp]
\centering
\includegraphics[width=0.9\textwidth]{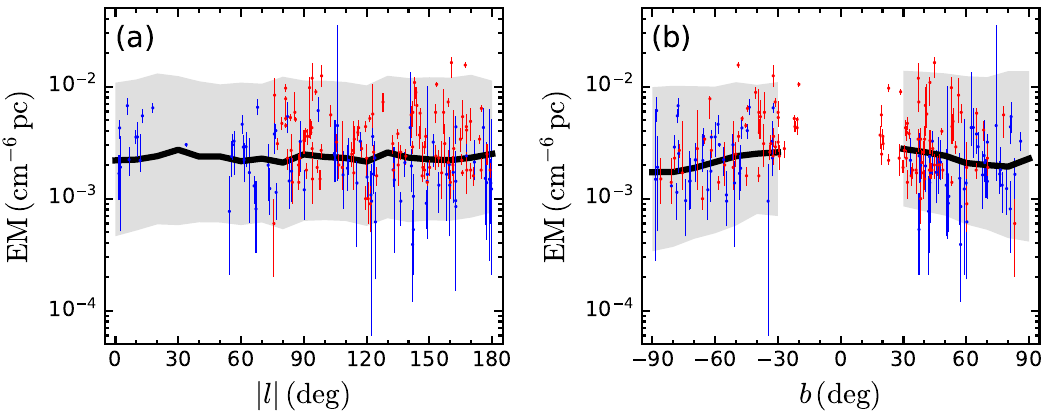}
\caption{Emission measure as a function of sky position. \textbf{(a)} EM versus absolute Galactic longitude $|l|$, the angular separation from the Galactic Center. \textbf{(b)} EM versus Galactic latitude $b$. The black line shows the median for all MW-mass galaxies from TNG50, with the gray band representing the $16^{\rm th}$--$84^{\rm th}$ percentile range. Observational measurements toward the MW halo from {\it XMM-Newton} \citep[blue;][]{2013ApJ...773...92H} and {\it Suzaku} \citep[red;][]{2018ApJ...862...34N} are overlaid.
\label{fig:em_lb}}
\end{figure*}

To further examine the spatial structure of the hot gas, Figure~\ref{fig:em_lb} 
shows EM as a function of Galactic coordinates. Panel (a) displays EM versus absolute Galactic longitude $|l|$. The simulated values are broadly consistent with the  observational data, and no strong trend of EM with $|l|$ is found within this high-latitude sample ($|b|>30^\circ$). This indicates that the density-squared integral of the hot gas does not show a pronounced large-scale asymmetry toward the Galactic Center direction at these latitudes.  
Panel (b) shows EM versus Galactic latitude $b$. Both the simulations and the observations show a mild decrease in EM with increasing $|b|$, indicative of a vertically stratified density distribution in the hot corona.

%%%%%%%%%%%%%%%%%%%%%%%%%%%%%%%%%%%%%%%%%%%%%%%%%%%%%%%%%%% 
\subsection{\ion{O}{7} and \ion{O}{8} Absorption}
\label{sec:O7}

\begin{figure*}[tbp]
\centering
\includegraphics[width=0.9\textwidth]{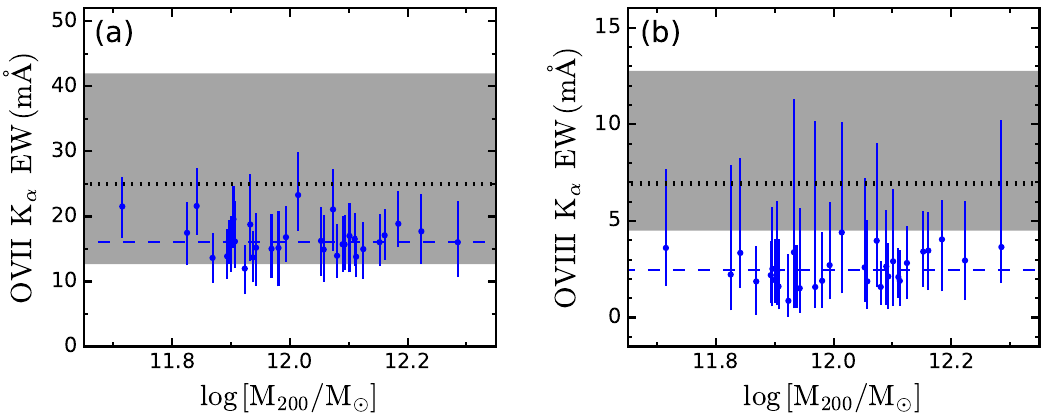}
\caption{Equivalent widths of the \ion{O}{7} and \ion{O}{8} K$\alpha$ absorption lines for MW-mass galaxies from TNG50, plotted as blue squares against halo mass $M_{200}$. Error bars represent the $1\sigma$ scatter from multiple sightlines per galaxy. The grey shaded bands show the observed $16^{\rm th}$--$84^{\rm th}$ percentile range of Galactic \ion{O}{7} \citep[left; ][]{2015ApJS..217...21F} and \ion{O}{8} absorption (right; \citealt{2012ApJ...756L...8G, 2010ApJ...717...74Z, 2014ApJ...785L..24F, 2016MNRAS.457.4236B, 2019ApJ...882L..23D}). Blue dashed and grey dotted lines indicate the respective median values.
\label{fig-O78}}
\end{figure*}

The strengths of the \ion{O}{7} and \ion{O}{8} absorption lines provide key constraints on the temperature structure of the hot halo. Figure~\ref{fig-O78} compares the simulated equivalent widths with the observed distributions for the Galactic halo.
For the \ion{O}{7} line, the majority of simulated galaxies falls within the observed scatter. The median simulated EW is approximately 36\% lower than the observational median, but remains consistent with the observational dispersion. In contrast, the simulated \ion{O}{8} absorption shows a clear and significant deficit: the median EW is about 65\% lower than observed, residing outside the observational range, with nearly all simulated galaxies lying below it.

The differing agreement reflects the distinct temperature sensitivities of the two ions. \ion{O}{7} is abundant over a broad temperature range ($T \sim 3\times10^5 - 2\times10^6$\,K), while \ion{O}{8} forms predominantly in a narrower, hotter window ($T \sim 1.6-3.2\times10^6$\,K), with its peak abundance near $T\approx2.5\times10^6$\,K under collisional ionization equilibrium \citep[e.g.,][]{2001ApJ...556L..91S}. Therefore, the significant deficit in \ion{O}{8} absorption indicates a shortage of gas in the hotter phase traced by this ion within the simulated halos. 

%%%%%%%%%%%%%%%%%%%%%%%%%%%%%%%%%%%%%%%%%%%%%%%%%%%%%%%%%%% 
\section{Discussion} 
\label{sec:discuss}

Our analysis presents a detailed comparison between the hot halos produced by TNG50 and observations, revealing both successes and clear limitations. The simulation reproduces key global and inner-halo observables for MW-mass galaxies: the soft X-ray luminosity, inner-region surface brightness, and overall emission measure are consistent with observations \citep{2013ApJ...773...92H, 2016MNRAS.457.1385W}. 
The good match in EM indicates that TNG50 generates approximately the correct density-squared integral (and thus hot gas mass) along inner-halo sightlines. Furthermore, the strength of \ion{O}{7} absorption, a tracer of warm-hot gas at $T \sim 10^6$\,K, is largely consistent with observational data \citep{2015ApJS..217...21F}. These agreements suggest that the simulation captures key aspects of the hot gaseous halo, particularly its mass and the characteristic temperature of its X-ray-bright component.

However, our analysis identifies two significant and interconnected discrepancies that point to specific limitations in the current feedback implementation of TNG50. First, the simulated \ion{O}{8} absorption is systematically weaker than observed (Figure~\ref{fig-O78}), indicating a pronounced deficit of gas in the temperature range $T \sim (1.6-3.2) \times 10^6$\,K \citep{1998A&AS..133..403M}. 
Second, and likely related, the simulated X-ray surface brightness profile is fainter and declines more steeply with radius than the SDSS stacking results \citep{2024A&A...690A.267Z}. This reveals that the hot gas in TNG50 is too centrally concentrated and lacks a sufficiently extended, diffuse component, at least relative to that population.

Several factors may contribute to these discrepancies. One possibility concerns the relative roles of different AGN feedback modes. The radiative ``quasar mode'' feedback, which dominates during earlier evolutionary stages and in more actively accreting systems, may be insufficient to establish an extended hot corona. For many MW-mass galaxies, the kinetic ``radio mode'' may only become active recently or intermittently, leaving limited time for its redistributive effects to operate. Even when active, the kinetic feedback may deposit energy too violently. 
Rather than gently lifting gas outward through outflows or buoyant bubbles, the energy injection could overheat gas in situ, pushing it to temperatures beyond the \ion{O}{8}-forming window ($T \gtrsim 3 \times 10^6$\,K) while failing to redistribute it to larger radii \citep{2021MNRAS.508.4667P}. This interpretation is consistent with independent findings that the TNG model produces a hotter-than-observed CGM around MW analogues \citep{2020ApJ...893L..24O} and drives high-velocity outflows in these systems \citep{2023MNRAS.518.5754R}.

Other factors may also contribute to the apparent discrepancy with the SDSS sample. Limitations in the treatment of stellar feedback could affect the energy balance in the inner halo, and numerical resolution effects might influence the ability to resolve multiphase structure and mixing processes. Additionally, the simulated and observed samples have slightly different selection criteria: our TNG50 sample is restricted to $M_{200} < 10^{13}\,M_\odot$, whereas the SDSS stacking may include a small fraction of more massive halos, which could boost the stacked signal. The eROSITA stacking analyses also involve assumptions about background subtraction, contamination from unresolved sources, and the conversion from stacked signal to individual halo properties.

We note that the nearby $L^*$ galaxy sample studied by \citet{2026arXiv260116499H} exhibits a steeper X-ray surface brightness profile than the SDSS sample, both derived from eROSITA stacking analyses, and is broadly consistent with TNG50 predictions. 
\citet{2026arXiv260116499H} selected 474 galaxies within 50\,Mpc, each individually inspected, whereas the SDSS sample contains $\sim 30,000$ MW-mass central galaxies and may include a small fraction of more massive halos that boost the stacked signal at large radii.
The consistency between the nearby $L^*$ sample and TNG50 suggests that part of the tension with the SDSS stacking results may be related to sample selection, distance, or contamination effects, and highlights the need for deeper observations of individual MW-mass galaxies to provide decisive validation.

In contrast to the SMUGGLE simulation, an isolated galaxy model without cosmological accretion, AGN feedback, or a pre-existing hot corona, TNG50 self-consistently incorporates these processes and successfully produces hot, gaseous halos around MW-mass galaxies. 
As a result, the SMUGGLE X-ray emission is confined to the inner regions \citepalias{2025ApJ...991..170Z} and falls below the TNG50 predictions at all radii (Figure~\ref{fig-sxp}), dropping below the TNG50 scatter beyond $\sim 70$\,kpc. However, the specific discrepancies identified here, namely the overly compact X-ray emission and the deficit of \ion{O}{8} absorption, indicate that even with these additional physical ingredients, the current feedback implementation in TNG50 still does not fully reproduce the observed hot halo properties.

These challenges are not unique to TNG50. 
For instance, \citet{2023MNRAS.525.1976T} analyzed soft X-ray emission line profiles (\ion{O}{7}, \ion{O}{8}, \ion{Fe}{17}) in TNG100, EAGLE, and SIMBA simulations at z=0, finding significantly different radial profiles among the three simulations with TNG100 predicting centrally concentrated emission consistent with our TNG50 broadband results. 
\citet{2025ApJ...993..125S} analyzed 28 MW-mass galaxies from seven simulation suites, eight of which come from TNG50, and found that TNG50 produces the shallowest broadband X-ray surface brightness profiles among all simulations studied, yet these remain steeper than eROSITA stacked observations.
\citet{2024ApJ...969...85S} similarly found that the detectability of hot CGM emission with future microcalorimeters depends sensitively on the underlying simulation model, with TNG and EAGLE halos exhibiting substantially different \ion{O}{7} and \ion{O}{8} surface brightness profiles and temperature structures. 
Taken together, these studies show that the discrepancy is not specific to TNG50 \citep[see also][]{2020MNRAS.499.5656R}. 
These comparisons demonstrate that the X-ray diagnostics presented here provide a valuable benchmark for testing feedback models.

%%%%%%%%%%%%%%%%%%%%%%%%%%%%%%%%%%%%%%%%%%%%%%%%%%%%%%%%%%%
\section{Summary} 
\label{sec:summary}

We have presented a systematic comparison between the hot gaseous halos of MW-mass galaxies in the TNG50 cosmological simulation and existing observational constraints. Our analysis is based on synthetic X-ray observables, including luminosity, surface brightness, and emission measure, together with \ion{O}{7} and \ion{O}{8} absorption line strengths. The main conclusions are as follows:

 \begin{enumerate}[label=(\roman*), align=left]

\item The simulation reproduces key global and inner-halo observables for MW-mass galaxies. The soft X-ray luminosities, inner-region surface brightness, emission measure, and \ion{O}{7} absorption strength all agree with MW measurements, indicating that TNG50 successfully generates the warm-hot ($T \sim 10^6$\,K) gas phase in the central regions.

\item The simulated hot halos are spatially too compact compared to the eROSITA stacking profile of SDSS galaxies. The azimuthally averaged X-ray surface brightness profile is broadly consistent with the SDSS data within $R \sim 50$\,kpc, but falls below by up to $\sim 1$ dex at $R \gtrsim 100$\,kpc, indicating a lack of diffuse X-ray bright gas at large galactocentric distances relative to this population.

\item The simulations underproduce gas in a specific hotter temperature phase. The strength of the \ion{O}{8} absorption line, which traces gas at $T \sim (1.6-3.2) \times 10^6$\,K, is systematically weaker than observed toward the Galactic halo, indicating a pronounced deficit of hotter-phase gas in the TNG50 halos.

 \end{enumerate}            

The TNG50 simulation reproduces the basic properties of hot gaseous halos, but the spatial distribution of the hot gas is too compact and the \ion{O}{8} absorption is systematically weaker than observed. These interconnected discrepancies indicate that the current implementation of AGN feedback still does not fully capture the observed properties of the hot halo. Resolving these tensions will require both improved AGN feedback prescriptions that distribute energy more effectively throughout the halo and deeper X-ray observations of individual MW-mass galaxies to distinguish between competing feedback models.

\begin{acknowledgements} 
This work is supported by the National SKA Program of China No. 2025SKA0150103, National Natural Science Foundation of China under Nos. 12550002, 12133008, 12221003, 11890692, 12233005. We acknowledge the science research grants from the China Manned Space Project with No. CMS-CSST-2021-A04 and No. CMS-CSST-2025-A10,
the Fujian Provincial Natural Science Foundation of China (Grant No. 2024J08001), and the Natural Science Foundation of Xiamen, China (No. 3502Z202472007). 
Q.Y. was supported by the European Research Council (ERC) under grant agreement no. 101040751.
\end{acknowledgements}

\software{ 
yt \citep{2011ApJS..192....9T}, 
Astropy \citep{2018AJ....156..123A}, 
Matplotlib \citep{2007CSE.....9...90H}, 
SciPy \citep{2020NatMe..17..261V}.
}

\bibliographystyle{aasjournal}
\bibliography{refer}{}
\end{CJK*}
\end{document}